\documentclass{llncs}

\usepackage{makeidx}  

\usepackage{graphicx}
\usepackage{verbatim}
\usepackage{amsmath}
\usepackage{algorithm}
\usepackage{algpseudocode}
\usepackage{algorithmicx}
\usepackage{tabularx}
\usepackage{multirow}
\usepackage{hhline}
\usepackage{color}

\usepackage[nolist]{acronym}
\begin{acronym}
	\acro{AAA}{Authentication, Authorization and Accounting}
	\acro{AP}{Access Point}
	\acro{API}{Application Programming Interface}
	\acro{BP}{Baseline Pseudonym}
	\acro{BS}{Base Station} 
    \acro{BSM}{Basic Safety Message}
    \acro{BYOD}{Bring Your Own Device}
	\acro{C2C-CC}{Car2Car Communication Consortium}
	\acro{CA}{Certification Authority}
	\acrodefplural{CA}{Certification Authorities}
	\acro{CN}{Common Name}
	\acro{CAM}{Cooperative Awareness Message}
	\acro{CIA}{Confidentiality, Integrity and Availability}
	\acro{CRL}{Certificate Revocation List}
	\acro{CDN}{Content Delivery Network}
	\acro{CSR}{Certificate Signing Requests}
	\acro{DAA}{Direct Anonymous Attestation}
	\acro{DDoS}{Distributed Denial of Service}
	\acro{DDH}{Decisional Diffie-Helman}
	\acro{DENM}{Decentralized Environmental Notification Message}
	\acro{DHT}{Distributed Hash Table}
	\acro{DoS}{Denial of Service}
	\acro{DPA}{Data Protection Agency}
	\acro{ECU}{Electronic Control Unit}
	\acro{ETSI}{European Telecommunications Standards Institute}
	\acro{ECDSA}{Elliptic Curve Digital Signature Algorithm}
	\acro{ECC}{Elliptic Curve Cryptography}
	\acro{EVITA}{E-safety Vehicle Intrusion protected Applications}
	\acro{FOT}{Field Operational Test}
	\acro{FPGA}{Field-Programmable Gate Array}
	\acro{GN}{GeoNetworking}
	\acro{GS}{Group Signatures}
	\acro{GM}{Group Manager}
	\acro{GBA}{Generic Bootstrapping Architecture}
	\acro{GPA}{Global Passive Adversary}
	\acro{GUI}{Graphic User Interface}
	\acro{HP}{Hybrid Pseudonym}
	\acro{HSM}{Hardware Security Module}
	\acro{HTTP}{Hypertext Transfer Protocol}
	\acro{IEEE}{Institute of Electrical and Electronics Engineers}
	\acro{IoT}{Internet of Things}
	\acro{ITS}{Intelligent Transport Systems}
	\acro{IT}{Information Technologies}
	\acro{IMSI}{International Mobile Subscriber Identity}
	\acro{IMEI}{International Mobile Station Equipment Identity}
	\acro{IdP}{Identity Provider}
	\acro{ISP}{Internet Service Provider}
	\acro{LEA}{Law Enforcement Agency}
	\acro{LTC}{Long-Term Certificate}
	\acro{LTCA}{Long-Term \acl{CA}}
	\acrodefplural{LTCA}{Long-Term \aclp{CA}}
	\acro{LDAP}{Lightweight Directory Access Protocol}
	\acro{LTE}{Long Term Evolution}
	\acro{LBS}{Location-based Service}
	\acro{MAC}{Message Authentication Code}
	\acro{MCA}{Message \ac{CA}}
	\acro{MEA}{Misbehavior Evaluation Authority}
	\acro{OBU}{On-board Unit}
	\acro{OCSP}{Online Certificate Status Protocol}
	\acro{OSN}{Online Social Network}
	\acro{PC}{Pseudonymous Certificate}
	\acro{PCA}{Pseudonymous \acl{CA}}
	\acrodefplural{PCA}{Pseudonymous \aclp{CA}}
	\acro{PDP}{Policy Decision Point}
	\acro{PEP}{Policy Enforcement Point}
	\acro{PIR}{Private Information Retrieval}
	
	\acro{PKI}{Public-Key Infrastructure}
	\acro{POI}{Point of Interest}
	\acro{PRECIOSA}{Privacy Enabled Capability in Co-operative Systems and Safety Applications}
	\acro{PRESERVE}{Preparing Secure Vehicle-to-X Communication Systems}
	\acro{P2P}{peer-to-peer}
	\acro{RA}{Resolution Authority}
	\acro{REST}{Representational State Transfer}
	\acro{RBAC}{Role Based Access Control}
	\acro{RCA}{Root \acl{CA}}
	\acro{RSU}{Roadside Unit}
	\acro{SAML}{Security Assertion Markup Language}
	\acro{SCORE@F}{Système COopératif Routier Expérimental Français}
	\acro{SeVeCom}{Secure Vehicle Communication}
	\acro{SIS}{Security Infrastructure Server}
	\acro{SP}{Service Provider}
	\acro{SSO}{Single-Sign-On}
	\acro{SoA}{Service-oriented-Approach}
	\acro{SOAP}{Simple Object Access Protocol}
	\acro{SAS}{Sample Aggregation Service}
	\acro{TS}{Task Service}
	\acro{TLS}{Transport Layer Security}
	\acro{TPM}{Trusted Platform Module}
	\acro{TTP}{Trusted Third Party}
	\acro{TVR}{Ticket Validation Repository}
	\acro{URI}{Uniform Resource Identifier}
	\acro{VANET}{Vehicular Ad-hoc Network}
	\acro{V2I}{Vehicle-to-Infrastructure}
	\acro{V2V}{Vehicle-to-Vehicle}
	\acro{V2X}{\ac{V2V} and/or \ac{V2I}}
	\acro{VC}{Vehicular Communication}
	\acro{VM}{Virtual Machine}
	\acro{VSS}{\ac{VC} Security Subsystem}
	\acro{WAVE}{Wireless Access in Vehicular Environments}
	\acro{WSDL}{Web Services Discovery Language}
	\acro{W3C}{World Wide Web Consortium}
	\acro{VANET}{Vehicular Ad-hoc Network}
	\acro{VPKI}{Vehicular Public-Key Infrastructure}
	\acro{WS}{Web Service}
	\acro{WoT}{Web of Trust}
	\acro{WSACA}{\ac{WAVE} Service Advertisement \ac{CA}}
	\acro{XML}{Extensible Markup Language}
	\acro{XACML}{eXtensible Access Control Markup Language}
	\acro{3G}{3rd Generation}
\end{acronym}

\begin{document}

	\frontmatter          
	\pagestyle{headings}  

	\mainmatter              

	\title{Resilient Collaborative Privacy for Location-Based Services}
	
	\author{Hongyu Jin and Panos Papadimitratos}

	\institute{Networked Systems Security Group\\
		KTH Royal Institute of Technology, Stockholm, Sweden\\
		\email{hongyuj@kth.se, papadim@kth.se}\\
		\texttt{www.ee.kth.se/nss}}
		
		\maketitle              
	
	\begin{abstract}
		\acp{LBS} provide valuable services, with convenient features for users. However, the information disclosed through each request harms user privacy. This is a concern particularly with \emph{honest-but-curious} \ac{LBS} servers, which could, by collecting requests, track users and infer additional sensitive user data. This is the motivation of both \emph{centralized} and \emph{decentralized} location privacy protection schemes for \acp{LBS}: anonymizing and obfuscating \ac{LBS} queries to not disclose exact information, while still getting useful responses. Decentralized schemes overcome the disadvantages of centralized schemes, eliminating anonymizers and enhancing users' control over sensitive information. However, an insecure decentralized system could pose even more serious security threats than privacy leakage. We address exactly this problem, by proposing security enhancements for mobile data sharing systems. We protect user privacy while preserving accountability of user activities, leveraging pseudonymous authentication with mainstream cryptography. Our design leverages architectures proposed for large scale mobile systems, while it incurs minimal changes to \ac{LBS} servers as it can be deployed in parallel to the \ac{LBS} servers. This further motivates the adoption of our design, in order to cater to the needs of privacy-sensitive users. We provide an analysis of security and privacy concerns and countermeasures, as well as a performance evaluation of basic protocol operations showing the practicality of our design.
	\end{abstract}
	
	\section{Introduction}
\label{sec:introduction}

The evolution and popularization of mobile Internet brings forth opportunities for service providers to cater to people's needs. \acfp{LBS} in particular respond to user queries based on their locations, available at their location-aware mobile devices. However, the improved relevance and precision of responses comes at a cost: users' privacy can be harmed~\cite{barkhuus2003location,myles2003preserving}; user location information can be used to reconstruct user trajectories and profile their activities or even infer their interests. In fact, the \ac{LBS} itself, i.e., its server(s), is uniquely positioned to undermine users' privacy, collecting rich information over time, for all locations a user (client or mobile device/application) submits queries from. Moreover, it can have a financial motivation to do so, seeking to push advertisements to users. As a result, increased concerns have been voiced and numerous efforts to safeguard user privacy led to a number of proposals, both \emph{centralized} and \emph{decentralized}.

%

Centralized schemes \cite{gedik2008protecting,mascetti2007spatial,mokbel2006new} introduce a new entity, an \emph{anonymizer}: it anonymizes a received client query (removing its identity attributes), obfuscates and/or blends the queries of multiple clients, and then sends them to the \ac{LBS} server(s). The location is obfuscated to a corresponding region with the client and at least $k-1$ other clients included, to achieve $k$-anonymity: the user is indistinguishable among these $k$ users. Clearly, these schemes are effective but this centralized approach seeks to solve the problem at hand based on the assumption that originally raised concerns for the \ac{LBS} servers;  they presume the anonymizer is trustworthy. But still, the anonymizer has all the rich information collected from client queries. If an LBS server can be curious and track or profile users, the question rises naturally: \emph{Why couldn't an anonymizer also breach the user privacy the same way}?

This challenge motivated a number of works that proposed decentralized privacy schemes. Similar, in spirit, to decentralized approaches overcoming disadvantages of centralized approaches for privacy-related problems in many areas \cite{mezzour2009privacy,han2011social,cutillo2009privacy}, \acp{LBS} user privacy protection can be achieved in a collaborative manner: without relying on an anonymizer. In particular, users can hide from the \ac{LBS} server by obtaining LBS-provided information from their neighbors~\cite{mobicrowd}.


Nonetheless, opening up the system functionality is a double-edged sword: it reduces the user exposure to the curious provider (\ac{LBS} or anonymizer) but it also exposes her to possibly faulty or misbehaving peers. In fact, risks in and abuses of, for example, \ac{P2P} systems~\cite{johnson2008evolution,kwok2002peer,zhou2005first} show that insecure decentralized schemes face serious problems. For example, users are threatened by exposure of their sensitive information to other peers or injected bogus data from malicious nodes. In~\cite{mobicrowd}, responses from the \ac{LBS} server are signed, thus they are self-verifiable while passed to other peers. However, this does not comply with many existing \ac{LBS} servers, which authenticate themselves and secure only pairwise communication over, e.g., a TLS channel, instead of signing the responses. In either case, peers could be uncooperative or offending.

This challenge exactly motivates our work, in the context of enhancing \ac{LBS} user privacy. The decentralized or \emph{collaborative} approach has clear advantages, enhancing the users' control over sensitive information: their exposure can be significantly reduced while they still obtain their sought quality of service (trading off mild delay for much better privacy). But this would be of no use if user peers could disrupt or even debilitate the collaborative querying part, by passing on bogus or irrelevant information, or excessively querying their peers. Even if \ac{LBS} responses were signed, still, misbehaving peers can aggressively consume resources of benign peers and obstruct the peer query-response operation. Or, worse even, abuse peer queries to also harm users' privacy based on the peer-to-peer data exchange.\footnote{Although, of course, an adversary would need a massive number of peers to collect, each one locally, the same information an \ac{LBS} would and is able to collect simply through its regular operations.}

This is what we address in this paper: we propose a security architecture for decentralized/collaborative privacy protection for \acp{LBS}. We are cognizant that already deployed \ac{LBS} servers would be unwilling to change their operations, thus we propose new components that are orthogonal to the \ac{LBS} servers and new functionality for the privacy-sensitive users. In fact, we conjecture that our scheme could even motivate \ac{LBS} servers to adopt and offer the collaborative privacy-enhancing scheme to their interested users. While, in turn, the users would be further motivated to embrace it knowing that it protects them from unwanted risks (manipulation, overloading) and, at the same time, safeguards their privacy. Moreover, we impose constraints on pseudonym usage to further protect users from being inundated with bogus data.


In the rest of the paper, we first outline requirements and discuss related work (Sec.~\ref{sec:problem}). Then, we present our proposed scheme (Sec.~\ref{sec:scheme}), we analyze the achieved security and privacy protection (Sec.~\ref{sec:analysis}), benchmark our protocol on mainstream mobile devices (clients) and provide a performance evaluation (Sec. \ref{sec:evaluation}), and conclude with our next steps (Sec.~\ref{sec:conclusion}).

	\section{Problem Statement and Related Work}
\label{sec:problem}


\subsection{System and Adversary Model}
\label{sec:model}

\begin{figure}[h!]
	\centering
	\includegraphics[width=0.6\columnwidth]{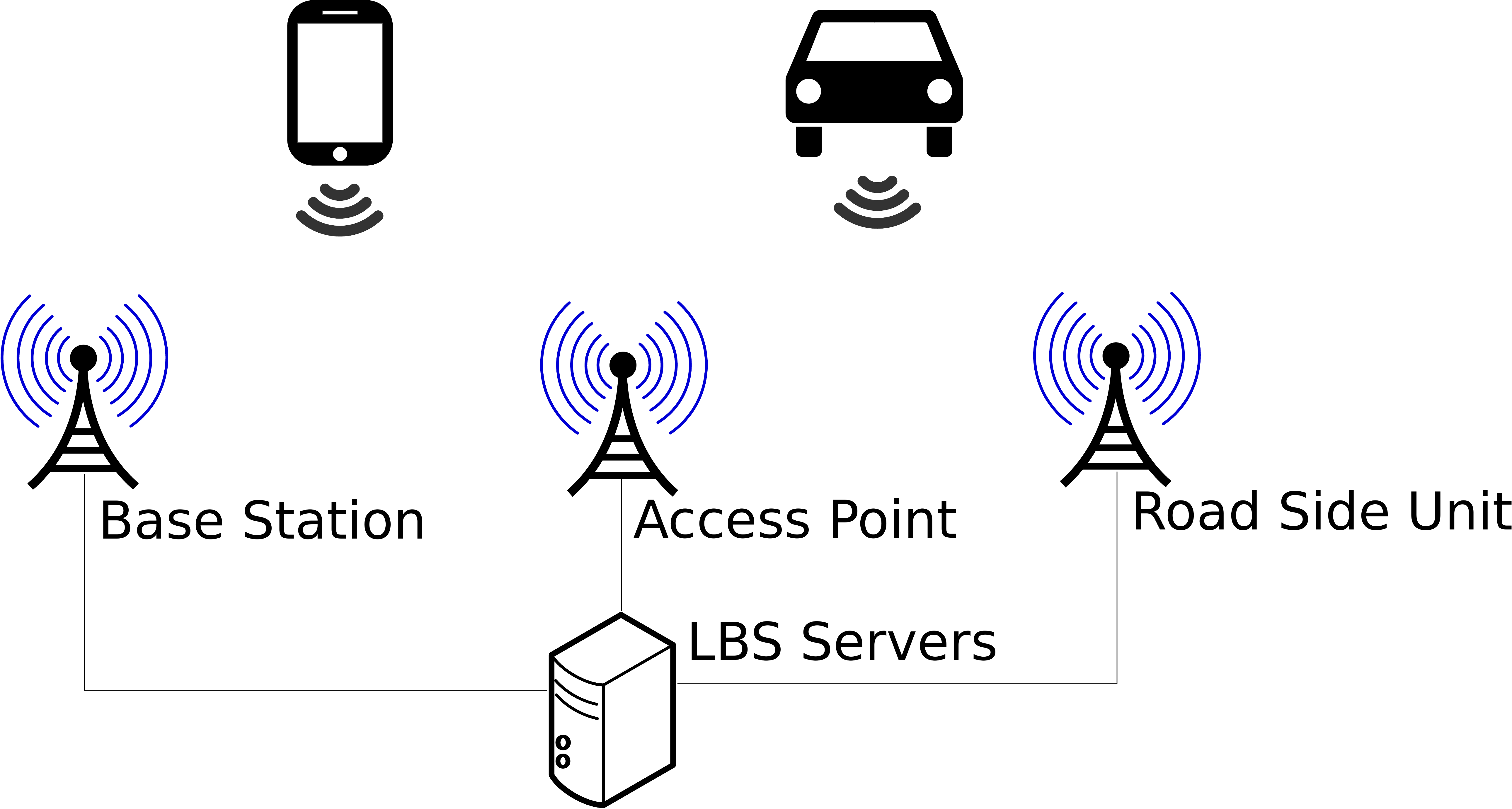}
	\caption{System Model}
	\label{fig:system}
\end{figure}

\textbf{System:} Consider a general model, as illustrated in Fig. \ref{fig:system}: Users' mobile clients, termed \emph{nodes} in the rest of the paper, can be smartphones, tablet PCs, \acp{OBU} in vehicles, etc. They are connected to the Internet through different channels (Wi-Fi and/or cellular network) and they are interested in different types of location-dependent information; e.g., specific \ac{POI} information, traffic status, environmental conditions, etc. Nodes are able to request these information from \ac{LBS} servers through the Internet. Nodes can in addition communicate with other nodes through ad-hoc connectivities, including Wi-Fi ad-hoc network, Wi-Fi Direct \cite{wifidirect}, LTE Direct \cite{ltedirect} and Bluetooth. This allows them to exchange information with other nodes; thus sharing information with each other or aggregating data obtained from multiple peers. We assume this is an alternative way of obtaining location-dependent information while hiding from the \ac{LBS} servers. This can be achieved by running an application on mobile devices; which obtains information from \ac{LBS} servers through provided APIs~\cite{googlemap,uber}, and shares the information with other nodes.

\textbf{Adversaries:} We assume that \ac{LBS} servers are \textit{honest-but-curious}: they follow the protocols, responding faithfully to their users' (nodes') queries. But they can trace the nodes (linking their queries), profile the nodes (recording their queries), and even deanonymize the nodes (inferring home and work sites). Such inferred sensitive data, based on the collected queries, could be commercially exploited. We maintain the same assumption for any third party, including the ones we introduce in our architecture (see Sec.~\ref{sec:scheme}).


Nodes can be honest, honest-but-curious or malicious. In the latter case, they can deviate from the collaborative protocol functionalities and policies, attacking the systems, notably their peer nodes: forging or tampering with responses, masquerading other nodes, excessively posting queries to their peers, seeking to exhaust their resources. The result could be degradation of the service users receive (i.e., being misled) or even a \ac{DoS} on the collaborative exchange. These would not only affect quality of service but also force honest nodes to expose themselves to the \ac{LBS} servers.


\subsection{Security and Privacy Protection Requirements}

We seek to thwart the aforementioned node misbehavior, while maintaining the benefit of ``hiding'' from the \ac{LBS} servers and obtaining useful information. 

\textbf{Authentication and Integrity}: Node messages, queries and responses, should allow their receiver to authenticate their sender and verify they were not modified or replayed from a previous exchange. 
We do not require strict identification of the sender (querier or responder) but at least validation that the sender is a legitimate participant of the \ac{P2P} operation.

\textbf{Non-repudiation and accountability}: The sender of any message (any action, in general) cannot deny having sent the message (taken the action). Any node can be tied to its actions, if need arises, and held accountable. Accordingly, it should be possible to have such nodes evicted from the system (the \ac{P2P} operation).

\textbf{Anonymity/Pseudonymity and unlinkability}: Nodes should not be identifiable, based on their \ac{P2P} interactions, and have their messages linked to their identities. Anonymity should be conditional, allowing the system to identify a misbehaving node (and evict it). Ideally, we want to make it impossible for any observer to link any two messages (e.g., queries) to the same node. But, for practical operation and efficiency/lower cost reasons, we require that node actions (messages) can be linked at most over a protocol selectable period $\tau$. Accordingly, any node can maintain one temporary identifier, a \emph{pseudonym}, for that same period. 

\textbf{Confidentiality} (optionally): The \ac{LBS}-originating content should be accessible only by legitimately participating nodes, possibly registered with the \ac{LBS} and the system/application that enables collaborative privacy protection.


%
%
%
%
%

\subsection{Related Work}

We discuss briefly decentralized approaches to enhance privacy for \acp{LBS}. As with centralized approaches, many decentralized schemes seek to provide $k$-anonymity: \ac{P2P} spatial cloaking \cite{chow2006peer} and MobiHide~\cite{ghinita2007mobihide} achieve $k$-anonymity by finding $k-1$ neighboring peers within a cloaked region. We do not dwell on how effectively this can be done (e.g., unlinkability under the assumption that nodes are not likely to move in the same direction). However, we note that the trust model among nodes (users) was not considered \cite{chow2006peer}; while MobiHide~\cite{ghinita2007mobihide} relies on a central server who maintains a list of active nodes and supports them on joining the clusters, which is a privacy threat for users. Along the same lines, AMOEBA \cite{amoeba07} protects user privacy by forming groups and delegating \ac{LBS} queries to group leaders, proposed for vehicular communication systems; the predictable mobility helps in that case. The formation of groups, of course, imposes additional complexity and overhead. If the conditions allow such group operation and it is effective, it could be beneficial. But such group formation and provision of $k$-anonymity is orthogonal to our work here, and it could possibly co-exist with (and even be facilitated by) our scheme, by explicitly addressing trust assumptions and providing a security architecture. 


Passing/sharing self-verifiable information among users helps to provide authentication and integrity \cite{mobicrowd}. Nodes cache information received from the \ac{LBS} server and pass to its neighbors when requested, thus decreasing exposure to the \ac{LBS} server. This is the approach we extend in this paper. It assumes that responses signed by the \ac{LBS} server are self-verifiable (manipulation by a node will be detected). However, even with such signatures, assuming of course a change on the side of \ac{LBS} (possibly considered unrealistic by some providers), a misbehaving node passing on tampered responses would remain ``invisible'' and continue attacking the system. This is exactly where this work comes in, protecting the system against node misbehavior. Moreover, it is interesting that this does not comply with many existing \ac{LBS} servers: user-/node-server communication is authenticated (and kept confidential) through end-to-end security (a secure channel, e.g., TLS, with the \ac{LBS} server), without signed responses.

The EU PRIME project~\cite{prime} proposes the use of anonymous credentials in the context of \ac{LBS}. This is related to our work, but a location intermediary is assumed between the \ac{LBS} and the mobile operator. The use of non-traditional public key cryptographic protocols has also been considered in~\cite{martucci2008self,calandriello2007efficient,calandriello2011performance,papadimitratos2008impact}, with special care for sybil-free operations, in spite of the relatively higher overhead for those cryptographic primitives.


	\section{Our Scheme}
\label{sec:scheme}

\subsection{Overview}
\label{sec:detail}

We assume basic collaborative functionality for nodes, sharing location-dependent information~\cite{mobicrowd}: before querying the \ac{LBS}, a node queries its neighbors/peers, who respond if they can; if no appropriate response is obtained in this \ac{P2P} manner, then the querier cannot but query the \ac{LBS}. To secure such a system, as per the requirements in Sec.~\ref{sec:problem}, we propose a security architecture and augment the basic \ac{P2P} functionality, and the node-to-\ac{LBS} communication. We assume nodes keep track of all communication in the vicinity, and react appropriately to P2P queries; listening if a query was already served by other nodes. This is straightforward to support in commodity wireless networks (e.g., Wi-Fi). Table \ref{table:notation} summarizes the used notation.

\begin{table}[htp]
	\caption{Notation}
	\centering
	\renewcommand{\arraystretch}{1.10}
	\begin{tabular}{l | *{1}{c} r}
		\hline \hline
		$LTCA$ & \emph{Long-Term Certification Authority} \\\hline
		$Lk/LK$ & \emph{Long-term Private/Public Key} \\\hline
		$LTC$  & \emph{\acl{LTC}} \\\hline
		$PCA$ & \emph{Pseudonymous Certification Authority} \\\hline
		$Sk/SK$ & \emph{Short-term Private/Public Key} \\\hline
		$PC$ & \emph{Short-term (Pseudonymous) Certificate} \\\hline
		$\{msg\}_\sigma$ & \emph{Signed msg} \\\hline
		$type_{_{poi}}$ & \emph{Type of POI} \\\hline
		$id$ & \emph{Node id or Query id} \\\hline
		$t/t_{now}$ & \emph{Timestamp/A fresh timestamp indicating current time} \\\hline
		$T_{timeout}$ & \emph{Timeout for peer response reception} \\\hline
		$SN$ & \emph{Serial Number} \\\hline
		$N$ & \emph{Number of needed responses to a peer query} \\\hline
		\hline
	\end{tabular}
	\renewcommand{\arraystretch}{1}
	\label{table:notation}
\end{table}


\begin{figure}[htp]
	\centering
	\includegraphics[width=0.7\columnwidth]{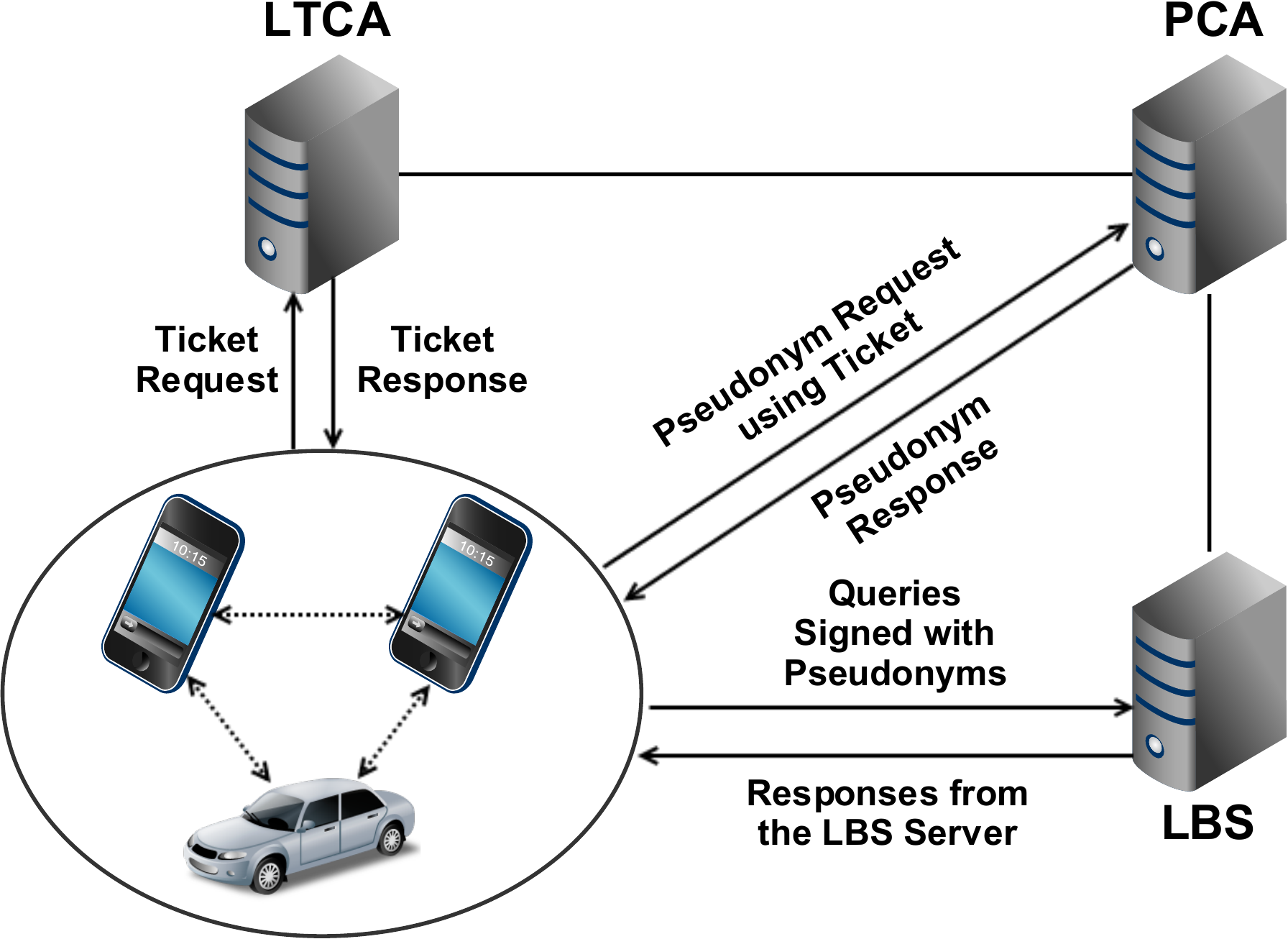}
	\caption{System Architecture}
	\label{fig:vespa_mobicrowd}
\end{figure}

Fig.~\ref{fig:vespa_mobicrowd} illustrates the proposed system architecture. We mandate that nodes are registered with an \emph{identity and credential management facility} that equips them with  \emph{short-lived anonymized credentials}. To do so, we require the nodes be registered with an \ac{LTCA} that maintains their long-term identities and issues \acp{LTC} for them. With the \ac{LTC}, a node obtains a ticket from the \ac{LTCA} and present the ticket to the \ac{PCA} to obtain \emph{\acp{PC}/pseudonyms}. The ticket is authenticated by the \ac{LTCA} but \emph{anonymized}: it does not reveal real identity of the node to the \ac{PCA}. Therefore, neither the \ac{LTCA} nor the \ac{PCA} can link the real identity of the node to the issued pseudonyms (thus, the messages signed under the pseudonyms). The details of the operation are presented next (Sec. \ref{sec:protocls}, \ref{sec:opt}).

We require that the pseudonyms be used to authenticate (with the corresponding cryptographic private key) \ac{P2P} queries and responses. They can be optionally used to authenticate queries to the \ac{LBS} (if its functionality allows that). The pseudonyms attest to the \emph{legitimate participation} of the node in node-to-\ac{LBS} or P2P communication. Furthermore, to prevent abuse of the node anonymity, our scheme provides \emph{conditional anonymity} and allows \emph{revocation of anonymity and eviction}. The node interactions with the facility entities are explained below. Moreover, we enforce ticket and pseudonym (lifetime) policies on the \acp{CA} and nodes, so that user privacy is protected to the full extent and the nodes have \emph{regulated access} to \ac{P2P} part of the \ac{LBS}. Finally, our extension of the \ac{P2P} functionality allows for increasing resilience and user-control: the querying node can seek multiple responses and can regulate the maximum rate at which it responds to queries.

\subsection{Protocols}\label{sec:protocls}

\textbf{Registration}: All nodes register with an \ac{LTCA}, which essentially acts as an identity provider. (1) A node generates a pair of long-term public/private keys, $LK$ and $Lk$, and (2) submits a \ac{CSR} (a self-signed $LK$ and other relevant information) to the \ac{LTCA}; (3) the node is issued with an \ac{LTC}. The whole exchange is secured with a TLS channel or done offline.

\begin{align}
C &: Lk, LK \\
C \rightarrow LTCA &: \{id_{_C},LK,others\}_{\sigma_{_{Lk}}} \\
LTCA \rightarrow C &: LTC_C= \{SN_{_{LTC}}, id_{_C}, LK, others\}_{\sigma_{_{LTCA}}}
\end{align}

\textbf{Ticket and Pseudonym Acquisition}: (4) A node requests a ticket with a desired pseudonym validity starting time, $t_{start}$. The length of ticket validity period is defined by system policy, thus no need to be specified by the node. (5) The \ac{LTCA} checks if a ticket was issued with an overlapping lifetime; if not, (6) it issues a ticket with validity period $[t_{start}^{'},t_{end}^{'}]$. The ticket validity period is computed by the \ac{LTCA} based on $t_{start}$ and the policy defined in~\cite{khod1412deploying} to prevent ticket and pseudonym linkability.

With the ticket in hand, the node can obtain a set of pseudonyms (7-9) from any associated \acp{PCA}, which acts as a service provider itself. There can be multiple that recognize/accept the \ac{LTCA} tickets; for the sake of presentation, without loss of generality, we refer to a single \ac{PCA}. The anonymized ticket does not reveal anything about the identity of the node (and the user) to the \ac{PCA}. This separation of duties concept is based on the work done in the context of vehicular communication systems~\cite{gisdakis2013serosa,khod1412deploying}. Both ticket and pseudonym acquisitions are protected with TLS channels. The ticket request is protected by mutual authentication; while the pseudonym request is protected by uni-directional (PCA-only) authentication, since the node is authenticated with the presented ticket.

	\begin{align}
	C \rightarrow LTCA &: ticket\_req\{t_{start}\}_{\sigma_{_C}} \\
	LTCA &: check(id_{_C}, t_{start}) \\
	LTCA \rightarrow C &: ticket = \{SN_{ticket},t_{start}^{'},t_{end}^{'}\}_{\sigma_{_{LTCA}}} \\
	C &: Sk, SK \\
	C \rightarrow PCA &: pseudonym\_req\{ticket, \{SK\}_{\sigma_{_{Sk}}}\}\\
	PCA \rightarrow C &: PC = \{SN_{pc},SK, t_{start}^{'}, t_{end}^{'}\}_{\sigma_{_{PCA}}}
	\end{align}


\begin{algorithm}[h]
	\caption{Querying thread (of a node)}
	\label{alg1}
	\begin{algorithmic}[1]
		\State Possesses a valid $PC$
		\State $query = \{loc, type_{\!_{poi}}\}$ \label{gen_q}
		
		\State $resp_{\!_{local}}=search(query)$
		\If {$resp_{\!_{local}}$ is satisfactory}
		\State $resp_{\!_{final}}=resp_{\!_{local}}$
		\Else
		\State $QUERY = \{id_q,t_{now}, query\}_{\sigma_{_{PC}}}$ \label{sign_q}
		\State $broadcast(\{QUERY, PC\})$
		\State Let $resp_{\!_{final}}=\phi$, $n = 0$, $t = t_{now} + T_{timeout}$
		\While {$\{RESP_i,PC_i\} = receiveRespBefore(t)$ and $n < N$} \label{receive}
		\State $RESP_i=\{id_q,t_{now}, resp\}_{\sigma_{_{PC_i}}}$
		\If {$verify(PC_i, LTC_{_{PCA}})$ and $verify(RESP_i,PC_i)$}
		\State $resp_{\!_{final}}=combine(resp_{\!_{final}}, resp_i)$
		\State $n = n + 1$
		\EndIf
		\EndWhile
		\If {$resp_{\!_{final}}$ is  not satisfactory}
		\label{lbs_start}
		\State $resp_{\!_{final}}= queryLBS(QUERY)$
		\EndIf \label{lbs_end}
		\State $cache(resp_{\!_{final}})$
		\EndIf
		\State \Return $resp_{\!_{final}}$
	\end{algorithmic}
\end{algorithm}

\begin{align}
Pr_{req}(i,j) =
\begin{cases}
1 &\text{i = j}\\
0 &\text{i $\neq$ j}
\end{cases} \label{eq:prob_req}
\end{align}

\begin{align}
Pr_{req}(i,j) = \frac{w_j * e^{-d_E(l_i, l_j)}}{\sum\limits_{k \in I}{w_k * e^{-d_E(l_i, l_k)}}} \label{eq:prob_req_random}
\end{align}

\textbf{\ac{P2P} Query}: Algorithm \ref{alg1} illustrates the querying thread of a node. As stated above, nodes cache locally responses received from the \ac{LBS} server (and other peers). When \ac{POI} information is needed, the local cache is checked first. If there is no match, it generates a signed query. To ensure unlinkability after a change of pseudonym, the node can randomly reset its IP and MAC address. The node waits for and possibly receives responses from its neighbors. It can specify in the query that $N$ responses are required in total from its peers and assign a query id ($id_q$). It then verifies the responses and combine them to form a final response. Each receiver could overhear the responses to the same query while the query is queued, and serve the query only while less than $N$ responses are overheard from the network (see Sec. \ref{sec:opt} for detail). Moreover, a node could adapt to current CPU usage and battery amount remaining, the rate at which to serve peer queries for further reducing the overhead. We do not formulate what is a satisfactory response to a query, it can be specified in the preferences of the application or determined through UI (e.g., a button indicating the user wants to query the LBS server directly) after being presented the peer responses.

\textbf{\ac{LBS} Query}: Nodes query the \ac{LBS} server only when they have to, e.g., when the information obtained from their neighbors is not satisfactory. The information obtained from the \ac{LBS} server is the essential resource for supporting \ac{P2P} function of our scheme. Nodes send the signed queries to the \ac{LBS} server. Then, the responses from the \ac{LBS} server are cached by the nodes.

\textbf{\ac{P2P} Query Processing}: As shown in Algorithm \ref{alg2}, when a node receives a peer query, it first verifies the attached pseudonym and checks if the attached pseudonym has been over-used (i.e., received queries signed under the same pseudonym exceeds the query rate allowed for one pseudonym). If not, it verifies the query and searches its cache. If successful in finding matching information, it signs and sends the response to the sender. This, of course, depends on whether or not $N$ responses to the same query have been overheard from the network.

\begin{algorithm}[H]
	\caption{Serving thread (of a node)}
	\label{alg2}
	\begin{algorithmic}[1] 
		\State Possesses a valid $PC$
		\State $\{QUERY_i, PC_i\} = receiveQuery()$
		\State $QUERY_i = \{id_q,t_{now},query\}_{\sigma_{_{PC_i}}}$
		
		\If {$verify(PC_i,LTC_{_{PCA}})$ and $verify(QUERY_i,PC_i)$}
		\State $query = \{loc, type_{\!_{poi}}\}$
		\State $resp=search(query)$
		\If {$resp \neq \phi$}
		\State $RESP=\{id_q,t_{now},resp\}_{\sigma_{_{PC}}}$
		\State $send(i,RESP)$
		\EndIf
		\EndIf
	\end{algorithmic}
\end{algorithm}

\textbf{Reporting misbehavior}:
We address post-misbehavior processing while the misbehavior detection is out of scope of this paper. However, we note that some types of misbehavior are straightforward to detect and confirm. For example, the honest \ac{LBS} responses can help finding out which of the contradictory responses to a query is/are bogus information. When a misbehavior is detected by a node, (\ref{eq4:1}) it sends to the \ac{RA}, the messages related to the misbehavior with pseudonyms attached. In case (\ref{eq4:6}) the messages are proved to be related to a misbehavior case, (\ref{eq4:2}) it sends the pseudonym (or multiple pseudonyms) to the \ac{PCA}, and (\ref{eq4:3}) the \ac{PCA} derives the $SN_{ticket}$ of the ticket that had been used to issue the pseudonym. (\ref{eq4:4}-\ref{eq4:5}) With the help of the \ac{LTCA}, the misbehaving node is exposed (and possibly evicted from the system).

\vspace{-1em}

\begin{align}
C \rightarrow RA &: \{\{msg\}_{\sigma_{_{PC_i}}},PC_i\}_{\sigma_{_{PC}}} \label{eq4:1}\\
RA &: judge(msg) \label{eq4:6}\\
RA \rightarrow PCA &: PC_i \label{eq4:2}\\
PCA \rightarrow RA &: SN_{ticket} \label{eq4:3}\\
RA \rightarrow LTCA &: SN_{ticket} \label{eq4:4}\\
LTCA \rightarrow RA &: id_i \label{eq4:5}
\end{align}

\vspace{-0.5em}


\subsection{Optimizations for Query Processing}\label{sec:opt}

Processing a query requires searching the node cache and two signature verifications: one for the attached certificate (pseudonym) and one for the sender's signature. Similarly, the response validation requires two signature validations, and checking if the nonce and location matches that of the query. To reduce communication and processing overhead, we propose the following optimizations:
\begin{itemize}

\item \textbf{Optimization 1}: A node could sign multiple queries/responses under the same pseudonym, thus it may omit attaching it to some of those - thus reducing communication overhead. Accordingly, the receiving nodes need to validate the pseudonym of the said node only once throughout the pseudonym lifetime. Then, they cache validated pseudonyms and omit their verification for successive queries/responses  signed under cached pseudonyms \cite{calandriello2011performance}. This way, the processing overhead is reduced, as only one signature needs to be validated for peer queries and responses. This is important, as the \ac{PCA} key would in principle have high security level, thus relatively higher verification delay.

\item \textbf{Optimization 2}: Each node could opportunistically cache responses to popular queries (both overheard and locally generated), assuming such information is likely to become useful later. The popularity can be determined by occurrence frequency of type of \ac{POI} in queries or responses, while it is protocol selectable. In case an incentive scheme is used, caching popular responses would provide increased rewards.
	
\item \textbf{Optimization 3}: Requesting multiple, $N$, responses allows cross-checking (Sec. \ref{sec:analysis}), but could waste resources if responses are sent after the querier obtained all needed responses. Responders can back-off randomly and overhear communications, counting responses to the specific query, based on its $id_q$; then, at the end of the back-off they respond only if less than $N$ responses were overheard.

\end{itemize}

	\section{Security and Privacy Analysis}
\label{sec:analysis}

In this section, we explain how the security and privacy requirements are addressed and how malicious behavior is thwarted.

\textbf{Authentication, integrity, and confidentiality:} The communication of the nodes with the \acp{CA} and the \ac{LBS} server is carried over TLS channels, thus providing end-to-end security. Signing \ac{P2P} messages under pseudonyms provides authentication and integrity. While confidentiality of \ac{P2P} communication is optional, any two nodes can establish a shared session key and mutually authenticate each other leveraging their pseudonymous certificates, encrypting the response(s) with the session key. Thus, only users registered with the system have access to \ac{LBS}-provided information.

\textbf{Non-repudiation and accountability:} The use of public key cryptography and the digital signatures ensure non-repudiation. Any suspected misbehavior (messages deemed to be inconsistent, bogus, etc., by a node) can be linked to the signer's pseudonym. This can be reported to the security infrastructure and the \ac{LTCA} and \ac{PCA} can jointly identify the node and if necessary evict it - revoking valid pseudonyms and/or preventing it from obtaining new pseudonyms.



\textbf{Unlinkability:} Keeping ticket request records at the \ac{LTCA} prevents nodes from excessively requesting pseudonyms with overlapping lifetimes - we enforce non-overlapping pseudonym lifetimes and similarly to \cite{khod1412deploying} we enforce that all tickets and pseudonyms are issued at discrete times for all requests to the \ac{PCA}. This precludes linkability of pseudonyms of the same node based on time of issuance and lifetime - the likelihood is inversely proportional to the number of all active pseudonyms in the system. Actions of a node are linkable only as long as the same private key (under the same pseudonym) is used, that is, only over the period $\tau$. Setting this is a trade-off between unlinkability and efficiency. Changing node identifiers across the protocol stack (IP, MAC) precludes linkability across pseudonym changes.

\textbf{Node authentication and exposure to the \ac{LBS} server}: For subscriber-based \acp{LBS}, node authentication is necessary. Optionally, nodes can be authenticated to the \ac{LBS} server with long-term credentials or pseudonyms based on the requirements of a specific \ac{LBS} server. Use of long-term credentials would make the nodes identifiable and queries linkable. While pseudonymous authentication ensures nodes authentication without revealing their identities and breaching unlinkability, if done under different pseudonyms.

\textbf{Non-verifiable responses:} If the \ac{LBS} server signs responses, their integrity and timeliness can be readily verified. Otherwise, any malicious node could forge bogus responses, or any node create an arbitrary, valid yet unverifiable response based on its cache. We don't address this aspect here; we only suggest that queriers request redundant responses in order to cross-check them and infer valid information, e.g., extracting and using only information included in the majority of the responses from distinct peer nodes. Such a scheme warrants a separate investigation and it is part of future work. We note, however, that the authentication of the nodes and the constraints imposed by our scheme prevent an adversary from posing as multiple nodes and inundating the receiver with bogus responses.

Essentially, honest \ac{LBS} responses can serve as ground truth for detecting injected bogus data. Nodes can examine suspicious responses (e.g., contradictory responses from different peers) by querying the \ac{LBS} server, and consequently, downgrading and reporting deviant responders. Actually, a node dissatisfied by other responders trades off its exposure to the \ac{LBS} server for precise and genuine information, while at same time, it contributes to the common objective: decrease and balance exposure of nodes to the \ac{LBS} server.


\textbf{Thwarting clogging attacks:} An internal attacker could post a large amount of queries to fetch cached information from its neighbors and drain their resources. Limiting the number of received peer queries signed under a same pseudonym and enforcing non-overlapped pseudonym lifetimes address this problem. It ensures each node has only one valid pseudonym at any point; thus when the quota (with respect to one receiver) of currently valid pseudonym has been consumed completely, it will not have any more valid pseudonyms to generate queries. However, flooding with bogus pseudonyms or messages that attached with bogus signatures could consume a lot of client resources for verification, while they are not avoidable. This is the same even if \ac{LBS}-obtained information are signed. Attackers can still pass forged data to their neighbors to consume resources of benign nodes. As a remedy for our system, keys with relatively low security levels could be used. Considering the ephemeral nature of information transmitted in the system and short lifetimes of credentials; even if the keys are cracked, the attacker will no longer be interested in expired credentials by that time. The decision on key choice will be made based on cryptographic benchmarks (see Sec. \ref{sec:evaluation}).

\textbf{Exposure to the security infrastructure and collusion with the \ac{LBS}:} Though authorities we introduce are designed in a manner that protects the nodes from being traced, they could be honest-but-curious. However, any of the honest-but-curious \acp{LTCA} or \acp{PCA} cannot trace a user's actions (based on an eavesdropped transcript) - we refer to the analysis in \cite{khod1412deploying}. Moreover, if the the \ac{LBS} server authenticated nodes with pseudonyms, its collusion with the \ac{LTCA} would not reveal any information; collusion with the \ac{PCA} would only reveal the batch of pseudonyms obtained with the one presented by the \ac{LBS} server but not with past ones issued to the same node under a different ticket. Only the unlikely collusion of all three, the \ac{LBS} server, the \ac{LTCA} and the \ac{PCA}, would expose the user.


	\section{Performance Evaluation}
\label{sec:evaluation}

In this section, we demonstrate the practicality and applicability of our scheme. We show performance evaluation results for basic operations in our system with off-the-shelf components and popular platforms, i.e., RSA and ECDSA for public key cryptography and Android smartphone as user device. We find that a smartphone can easily handle high query rates from its neighbors, especially by using RSA keys with relatively low security levels.

\begin{table}[h!]
	\newcommand{\tabincell}[2]{\begin{tabular}{@{}#1@{}}#2\end{tabular}}
	\centering{\footnotesize
		\caption{Processing delay of cryptographic operations}
		\label{tab:co-1}
		\begin{tabular}{| c | c | c | c | c | c | c |}
			\hline
			\tabincell{c}{\textbf{\textit{Key Type}}} &  \tabincell{c}{\textbf{\textit{Security}}\\\textbf{\textit{Level}}\\(bits)} & \tabincell{c}{\textbf{\textit{Generation}}\\(ms)} & \tabincell{c}{\textbf{\textit{Sign}}\\(ms)} & \tabincell{c}{\textbf{\textit{Verify}}\\(ms)} & \tabincell{c}{\textbf{\textit{Signature}}\\\textbf{\textit{Size}}\\(bytes)}\\ \hline
			RSA-1024 & \textcolor{black}{80} & \textcolor{black}{400.86} & \textcolor{black}{4.63} & \textcolor{black}{0.78} & \textcolor{black}{128}\\  \hline
			RSA-2048 & 112 & 2104.59 & 21.18 & 1.21 & 256\\ \hline
			ECDSA-192 & 96 & 214.65 & 210.01 & 286.44 & 56\\ \hline
			ECDSA-224 & 112 & 251.66 & 251.91 & 345.95 & 63\\ \hline
		\end{tabular}}
	\end{table}

The most frequent and time consuming operations are signature generation and verification.\footnote{Compared to other domains \cite{khod1412deploying,gisdakis2014sppear}, which need frequent pseudonym changes to ensure unlinkability of messages transmitted in high rate, a relatively longer pseudonym lifetime is acceptable in our scheme, as one can expect relatively lower message rates. Thus, the performance is less affected by key generation operations.} Encryption of \ac{P2P} communication would incur also key establishment cost, thus some additional public key encryption, whose processing delay is on the same order of magnitude of signature verification delay. However, we do not explicitly consider it here, as it is optional. Table~\ref{tab:co-1} shows processing delays for cryptographic operations on a \emph{Sony Xperia Ultra Z} smartphone, which has a \emph{Quad-core 2.2 GHz Krait 400} CPU. We choose RSA and ECDSA algorithms, commonly used for public key cryptography. Note that ECDSA is standard in other applications, notably \acp{VANET} \cite{ieee16092}, due to its low key generation and signing delays and short signature sizes. However, the Spongy Castle library \cite{scprov}, the only library available for Android supporting ECDSA, is inefficient. We found that RSA key generation takes more time than ECDSA, but sign/verify operations take much less time than ECDSA. Actually, all the execution delays of ECDSA are abnormally high (compared to those in other libraries for other platforms). By checking the system logs of the smartphone, we found that for each cryptographic operation of ECDSA the application needs to free the heap 2 or 3 times. As a result, it increases significantly cryptographic latencies, due to the limited heap size of each application in Android and the high spatial overhead of ECDSA operations. Due to abnormalities of ECDSA operations in Android, RSA is preferred for our scheme.

	
\begin{table}[h!]
	\newcommand{\tabincell}[2]{\begin{tabular}{@{}#1@{}}#2\end{tabular}}
	\centering{\footnotesize
		\caption{Processing overhead for different operations}
		\label{tab:delay}
		\begin{tabular}{| c | c |}
				\hline
				\textbf{\textit{Operation}} & \textbf{\textit{Processing Overhead}}\\ \hline
				\textbf{\tabincell{c}{Message verification with\\cached pseudonym}} & \tabincell{c}{Message Verification}\\ \hline
				\textbf{\tabincell{c}{Message verification with\\non-cached pseudonym}}  & \tabincell{c}{Pseudonym Verification,\\Message Verification}\\ \hline
				\textbf{\tabincell{c}{Query generation}}  & \tabincell{c}{Message Signing}\\ \hline
				\textbf{\tabincell{c}{Response generation}}  & \tabincell{c}{Database Query,\\Message Signing}\\ \hline
		\end{tabular}}				
\end{table}

\begin{figure}[h!]
	\centering
	\includegraphics[width=0.75\columnwidth,keepaspectratio]{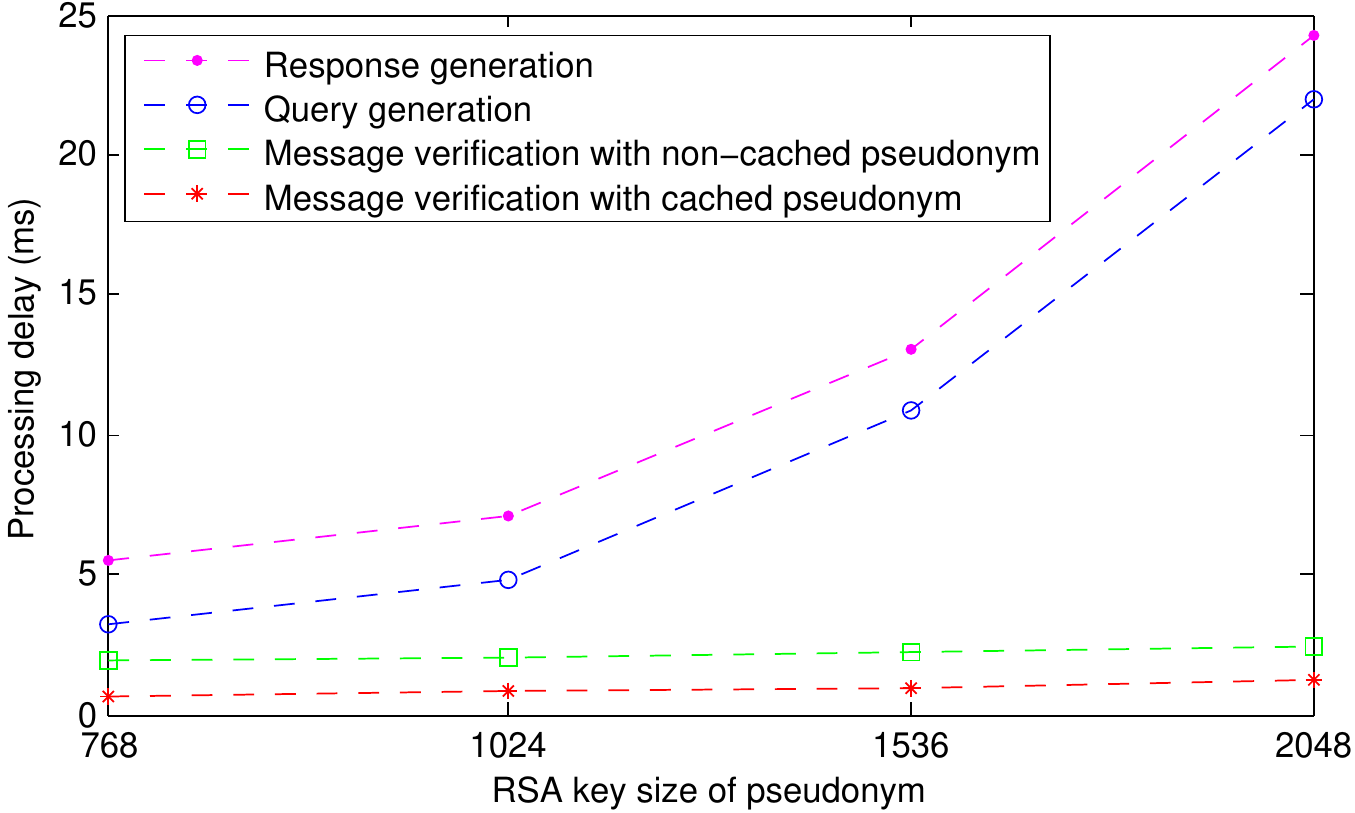}
	\caption{Processing delay under pseudonyms with different RSA key sizes, assuming an RSA-2048 certificate of the \ac{PCA}}
	\label{fig:delay}
\end{figure}

Table \ref{tab:delay} shows processing overhead for different operations and Fig. \ref{fig:delay} shows processing delays on the smartphone. Consider, for example, mobile phone user density in Spanish cities \cite{louail2014from}: Barcelona, the most densely populated in terms of mobile phone users in Spain, has around 3000 mobile phone users per $km^2$. Assuming Wi-Fi radio range of $100m$, there are around 100 peers within range. Assume all peers (e.g., in a landmark site or in the city center) need to query with a query rate per user equal to 1 $query/min$: this implies that a node would receive approximately 1.7 $queries/sec$. From Fig. \ref{fig:delay}, we can see that verifying a query, even with non-cached pseudonyms, would incur processing overhead for less than 3 $msec$, thus more than 300 $queries/sec$ could be verified. Peer response generation delay is highly dependent on local cache implementation and size of the cache. It includes a searching process and a signing operation. In our experiment, we use SQLite database. We assume 50 pieces of \acp{POI} are stored in the cache and 5 pieces of them matches each query. From Fig. \ref{fig:delay}, we see it takes approximately 7 $msec$ to generate a peer response with RSA-1024 key. This implies that a node could, if able and needing to, respond to 1.7 $queries/sec$ (the query rate we calculated earlier).
	
Based on availability of relevant information in the cache, only a part of them could be served; moreover, the actual latency could be significantly lower thanks to optimizations we proposed in Sec. \ref{sec:scheme}. More basically, a node can ``decide,'' based on, e.g., current CPU usage and battery amount remaining, whether or not to serve peer queries. A protocol-selectable parameter, the maximum rate at which to respond, can further reduce the overhead.

Peer query and response sizes are application and implementation dependent. In our experiments, we assume 25 byte queries. By encoding public keys and signatures into Base64 format and encapsulating into JSON format; the total size of a peer query, with an RSA-1024 pseudonym (public key and signature from the PCA) attached, is 980 bytes. The size of a peer response depends on how many information pieces it contains; while the signature and attached pseudonym sizes are same as those for a query. With most smartphones supporting IEEE 802.11 b/g/n with throughput between 11 and 300 Mbps, the above mentioned query rate would not incur heavy communication overhead. Clearly, receiving queries would affect posting own queries, as responses are received over the same channel. But, as mentioned above, beyond optimizations, a node can always stop serving queries beyond a threshold, to ensure it can obtain own needed information.

	\section{Conclusion}
\label{sec:conclusion}

We presented a decentralized secure and privacy protection scheme for \acp{LBS}. We leverage the concept of information sharing in \ac{P2P} systems for \ac{POI} information sharing, and further secure it in a privacy-preserving manner with pseudonym-based authentication. Through security and privacy analysis and performance evaluation, we show a system with high resiliency to different attacks and high practicality for the deployment. Our scheme can be extended in terms of optimizations we proposed. In our evaluation, we assume mobile nodes are evenly distributed. However, efficiency of our scheme in flash crowds needs to be evaluated with large scale simulation to show how peer queries for varying types of \ac{POI} information can be handled by load balancing among the nodes in the crowds. Moreover, with simulation, we can quantify user privacy and determine optimal parameters for the optimizations we proposed. An incentive scheme and cross-checking mechanism can be integrated to promote user participation and improve attack resiliency.

	\bibliographystyle{abbrv}
	\bibliography{references.bib}

\begin{thebibliography}{10}

\bibitem{googlemap}
Google maps api.
\newblock https://developers.google.com/maps/.

\bibitem{ltedirect}
Lte direct.
\newblock https://www.qualcomm.com/invention/technologies/lte/direct.

\bibitem{scprov}
{{T}he {S}pongy {C}astle {C}ryptography {API}s}.
\newblock https://rtyley.github.io/spongycastle/.

\bibitem{uber}
Uber api.
\newblock https://developer.uber.com/.

\bibitem{wifidirect}
Wi-fi direct.
\newblock https://rtyley.github.io/spongycastle/.

\bibitem{prime}
{PRIME Framework Version. 3}.
\newblock https://www.prime-project.eu/prime\_products/reports/fmwk/, 2008.

\bibitem{ieee16092}
{IEEE Standard for Wireless Access in Vehicular Environments Security Services
  for Applications and Management Messages}.
\newblock {\em IEEE Std 1609.2-2013}, 2013.

\bibitem{barkhuus2003location}
L.~Barkhuus and A.~K. Dey.
\newblock Location-based services for mobile telephony: a study of users'
  privacy concerns.
\newblock In {\em INTERACT}, Cape Town, South Africa, Sept. 2003.

\bibitem{calandriello2007efficient}
G.~Calandriello, P.~Papadimitratos, J.-P. Hubaux, and A.~Lioy.
\newblock Efficient and robust pseudonymous authentication in vanet.
\newblock In {\em ACM VANET}, Montreal, Canada, 2007.

\bibitem{calandriello2011performance}
G.~Calandriello, P.~Papadimitratos, J.-P. Hubaux, and A.~Lioy.
\newblock On the performance of secure vehicular communication systems.
\newblock {\em IEEE TDSC}, 2011.

\bibitem{chow2006peer}
C.-Y. Chow, M.~F. Mokbel, and X.~Liu.
\newblock A peer-to-peer spatial cloaking algorithm for anonymous
  location-based service.
\newblock In {\em ACM GIS}, New York, NY, Nov. 2006.

\bibitem{cutillo2009privacy}
L.~A. Cutillo, R.~Molva, and T.~Strufe.
\newblock Privacy preserving social networking through decentralization.
\newblock In {\em IEEE/IFIP WONS}, Snowbird, Utah, Feb. 2009.

\bibitem{gedik2008protecting}
B.~Gedik and L.~Liu.
\newblock Protecting location privacy with personalized k-anonymity:
  Architecture and algorithms.
\newblock {\em IEEE Transactions on Mobile Computing}, Jan 2008.

\bibitem{ghinita2007mobihide}
G.~Ghinita, P.~Kalnis, and S.~Skiadopoulos.
\newblock Mobihide: a mobilea peer-to-peer system for anonymous location-based
  queries.
\newblock In {\em SSTD}. Boston, MA, July 2007.

\bibitem{gisdakis2014sppear}
S.~Gisdakis, T.~Giannetsos, and P.~Papadimitratos.
\newblock Sppear: security \& privacy-preserving architecture for
  participatory-sensing applications.
\newblock In {\em ACM WiSec}, Oxford, UK, July 2014.

\bibitem{gisdakis2013serosa}
S.~Gisdakis, M.~Lagan{\`a}, T.~Giannetsos, and P.~Papadimitratos.
\newblock Serosa: Service oriented security architecture for vehicular
  communications.
\newblock In {\em IEEE VNC}, Boston, MA, Dec. 2013.

\bibitem{han2011social}
L.~Han, B.~Nath, L.~Iftode, and S.~Muthukrishnan.
\newblock Social butterfly: Social caches for distributed social networks.
\newblock In {\em PASSAT}, Boston, MA, Oct. 2011.

\bibitem{johnson2008evolution}
M.~Johnson, D.~McGuire, and N.~Willey.
\newblock The evolution of the peer-to-peer file sharing industry and the
  security risks for users.
\newblock In {\em HICSS}, Waikoloa, Big Island, Hawaii, Jan. 2008.

\bibitem{khod1412deploying}
M.~Khodaei, H.~Jin, and P.~Papadimitratos.
\newblock Towards deploying a scalable \& robust vehicular identity and
  credential management infrastructure.
\newblock In {\em IEEE VNC}, Paderborn, Germany, Dec. 2014.

\bibitem{kwok2002peer}
S.~H. Kwok, K.~R. Lang, and K.~Y. Tam.
\newblock Peer-to-peer technology business and service models: risks and
  opportunities.
\newblock {\em Electronic Markets}, 2002.

\bibitem{louail2014from}
T.~{Louail}, M.~{Lenormand}, O.~G. {Cantu Ros}, M.~{Picornell}, R.~{Herranz},
  E.~{Frias-Martinez}, J.~J. {Ramasco}, and M.~{Barthelemy}.
\newblock {From mobile phone data to the spatial structure of cities}.
\newblock {\em Scientific Reports}, June 2014.

\bibitem{martucci2008self}
L.~A. Martucci, M.~Kohlweiss, C.~Andersson, and A.~Panchenko.
\newblock Self-certified sybil-free pseudonyms.
\newblock In {\em ACM WiSec}, Alexandria, VA, 2008.

\bibitem{mascetti2007spatial}
S.~Mascetti, C.~Bettini, D.~Freni, and X.~S. Wang.
\newblock Spatial generalisation algorithms for lbs privacy preservation.
\newblock {\em Journal of Location Based Services}, 2007.

\bibitem{mezzour2009privacy}
G.~Mezzour, A.~Perrig, V.~Gligor, and P.~Papadimitratos.
\newblock Privacy-preserving relationship path discovery in social networks.
\newblock In {\em CANS}. Japan, Dec. 2009.

\bibitem{mokbel2006new}
M.~F. Mokbel, C.-Y. Chow, and W.~G. Aref.
\newblock The new casper: query processing for location services without
  compromising privacy.
\newblock In {\em Proceedings of the 32nd international conference on Very
  large data bases}, Seoul, Korea, Sept. 2006.

\bibitem{myles2003preserving}
G.~Myles, A.~Friday, and N.~Davies.
\newblock Preserving privacy in environments with location-based applications.
\newblock {\em IEEE Pervasive Computing}, 2003.

\bibitem{papadimitratos2008impact}
P.~Papadimitratos, G.~Calandriello, A.~Lioy, and J.-P. Hubaux.
\newblock {I}mpact of {V}ehicular {C}ommunication {S}ecurity on
  {T}ransportation {S}afety.
\newblock In {\em IEEE INFOCOM MOVE}, Phoenix, AZ, 2008.

\bibitem{amoeba07}
K.~Sampigethaya, M.~Li, L.~Huang, and R.~Poovendran.
\newblock Amoeba: Robust location privacy scheme for vanet.
\newblock {\em IEEE JSAC}, 2007.

\bibitem{mobicrowd}
R.~Shokri, G.~Theodorakopoulos, P.~Papadimitratos, E.~Kazemi, and J.-P. Hubaux.
\newblock Hiding in the mobile crowd: Location privacy through collaboration.
\newblock {\em IEEE TDSC}, 2014.

\bibitem{zhou2005first}
L.~Zhou, L.~Zhang, F.~McSherry, N.~Immorlica, M.~Costa, and S.~Chien.
\newblock A first look at peer-to-peer worms: threats and defenses.
\newblock In {\em Proceedings of the 4th International Conference on
  Peer-to-Peer Systems}. Konstanz, Germany, Aug. 2005.

\end{thebibliography}

\end{document}